\documentclass[Journal]{IEEEtran}

\usepackage{xspace,amsmath,amssymb,amsfonts,epsfig,syntonly}
\usepackage{cite,bm,color,url,textcomp,empheq,boxedminipage}
\usepackage{algorithmicx,algorithm,algpseudocode}
\usepackage{epstopdf}
\usepackage{empheq}
\usepackage{pifont}
\usepackage{graphicx,graphics,subfigure}  
\usepackage{multirow,multicol}
\usepackage{psfrag}    
\usepackage{stfloats}
\usepackage{url}

\usepackage{array}

\usepackage{xspace,amsmath,amssymb,epsfig,syntonly,cite}

\newtheorem{lemma}{\underline{Lemma}}[section]

\newtheorem{proposition}{\underline{Proposition}}[section]

\newtheorem{remark}{\underline{Remark}}[section]

\newcommand{\mv}[1]{\mbox{\boldmath{$ #1 $}}}

\DeclareMathOperator*{\argmin}{arg\,min}

\long\def\symbolfootnote[#1]#2{\begingroup
\def\thefootnote{\fnsymbol{footnote}}
\footnote[#1]{#2}\endgroup}

\psfull
\allowdisplaybreaks[4]

\usepackage[T1]{fontenc}
\usepackage{ifpdf}
 \ifpdf
 \else
 \fi

\begin{document}

\title{Decentralized Robust Transceiver Designs for MISO SWIPT Interference Channel}

\author{Feng Wang, Tao Peng, and Yongwei Huang

\thanks{This paper was presented in part at {\em Proc. IEEE VTC}, Nanjing, China, May 2016, pp. 1--5\cite{VTC16}.}
\thanks{F. Wang is with the school of Information Engineering, Guangdong University of Technology, Guangzhou 510006, China (e-mail: fengwang.nl@gmail.com).}
\thanks{T. Peng is with the Key Laboratory for Information Science of Electromagnetic Waves (MoE), the Department of Communication Science and Engineering, Fudan University, Shanghai 200433, China (e-mail: 13210720103@fudan.edu.cn).}
\thanks{Y. Huang is with the school of Information Engineering, Guangdong University of Technology, Guangzhou 510006, China. (e-mail: ywhuang@gdut.edu.cn, eeyw@ust.hk).}
}



\maketitle

\begin{abstract}
This paper considers a $K$-user multiple-input single-output (MISO) interference channels for simultaneous wireless information and power transfer (SWIPT), where each multi-antenna transmitter serves a single-antenna receiver per user pair. All receivers perform simultaneously information processing and energy harvesting (EH) based on the receive power-splitting (PS) architectures. Assuming imperfect channel state information (CSI) at the transmitters, we develop an optimal robust transceiver design scheme that minimizes the total transmission power under the worst-case signal-to-interference-plus-noise ratio (SINR) and energy harvesting (EH) constraints at the receivers, by jointly optimizing transmit beamforming and receive PS ratio per receiver. When the CSI uncertainties are bounded by ellipsoidal regions, it is shown that the worst-case SINR and EH constraints per receiver can be recast into quadratic matrix inequality forms. Leveraging semidefinite relaxation technique, the intended robust beamforming and PS (BFPS) problem can be relaxed as a tractable (centralized) semidefinite program (SDP). More importantly, relying on the state-of-the-art alternating direction method of multipliers (ADMM) in convex optimization, we propose a {\em decentralized} algorithm capable of computing the optimal robust BFPS scheme with local CSI and limited information exchange among the transmitters. It is shown the proposed decentralized algorithm is guaranteed to converge to the optimal centralized solution. Numerical results are provided to demonstrate the merits of the proposed approaches.
\end{abstract}




%

\section{Introduction}\label{sec:introduction}
By harvesting renewable energy from environmental sources (e.g., solar and wind), energy harvesting (EH) are expected to achieve energy self-sufficiency operation and reduce carbon footprint for low-power wireless devices such as sensor nodes and Internet-of-things (IoT) devices\cite{Ulukus15}. Due to the intermittent and randomness nature of the available energy, EH also brings new theoretical and design challenges to wireless communication \cite{Wang13,Chen15,Wang15,Chen16,Nan16}. In addition, emerging radio-frequency (RF) signal based wireless power transfer (WPT) provides a {\em controllable} solution for EH with sustainable energy supply\cite{Vis13}, by deploying dedicated energy transmitters.

With a joint design of WPT and wireless communication, simultaneous wireless information and power transfer (SWIPT) paradigm has been proposed to achieve ubiquitous wireless communications in a self-sustainable manner\cite{Zhang13,Var08,Kri14,Gro10}. In SWIPT, RF signals carry both information and energy. From an information-theoretic perspective, \cite{Var08,Gro10} addressed the fundamental capacity-energy tradeoff for SWIPT. Some practical design and signal processing issues for multi-input multi-output (MIMO) SWIPT systems were then examined in \cite{Zhang13,Kri14,Gro10}. As the simultaneous information processing and EH by one circuit is not possible with existing techniques, the practical receiver architectures are generally classified into two types, i.e., timing-switching (TS) and power-splitting (PS). For TS-based SWIPT systems, orthogonal time division is required for WPT and wireless communication such that the transmitter (Tx) sends the energy-bearing and the information-bearing signals orthogonally, while the receiver (Rx) respectively handles the operations of EH and information processing. For PS-based SWIPT systems, orthogonal time division is not required and the Rxs can split the received RF signal into two parts for EH and information processing in parallel; hence, the PS ratio is a key design variable for each Rx.

On the other hand, beamforming, an advanced signal processing technique of multi-antenna communications, has been proposed in various wireless communication systems to improve the information transmit rate and reliability \cite{Fang13,Feng17,Huang13,Chao12} and is also expected to improve the WPT efficiency in SWIPT\cite{Lu15,Zeng17,Fang15,Tim14,Xu14,Shi14tsp}. With beamforming, one can better exploit the spatial characteristics of the propagation channel for information/energy transmission. Beamforming design for SWIPT systems has received extensive attentions in the literature (see, e.g., \cite{Lu15,Zeng17} and references therein). The information and energy beamforming designs were investigated for multi-input single-out (MISO) downlinks \cite{Xu14}. For the PS-based architecture, jointly optimal beamforming and receive PS (BFPS) designs were pursued for MISO downlinks \cite{Tim14}, interference channels \cite{Shi14tsp}, and two-way relay channels \cite{Fang15}. The existing works in\cite{Xu14,Fang15,Tim14,Shi14tsp} assumed that perfect channel state information (CSI) is available at Txs. Note that perfect CSI is rarely available for practical SWIPT systems due to the channel estimation errors, latency feedback, and hardware impairments. In addition, a high signaling-overhead cost is usually required for high-quality CSI at the Tx. Therefore, robust designs by taking into account imperfect CSI are necessary and desirable for relieving the signaling-overhead burden while guaranteeing the users' quality-of-services (QoS) in communication and WPT \cite{Xiang12}. The robust SWIPT designs were also recently pursued for MISO downlinks \cite{Feng15}, interference channels \cite{Zhao15}, and secrecy communications \cite{Ng14}. Note that most of the existing literature aimed to obtaining centralized solutions.

Motivated by robust SWIPT designs with the PS receiver architecture, this paper considers a $K$-user MISO interference channel for SWIPT. We pursue a robust decentralized BFPS design based on the celebrated alternating direction method of multiplier (ADMM) in convex optimization \cite{Boyd08}. Our main contributions are summarized as follows.

\begin{enumerate}
\item Assuming the bounded CSI uncertainty at each Tx, we propose a robust BFPS design for the total transmission power minimization subject to the {\em worst-case} signal-to-interference-plus-noise ratio (SINR) and EH constraints at the Rxs. These two types of constraints guarantee the QoS for SWIPT users in information transmission reliability and EH reliability, respectively.

\item To circumvent the dilemma of the infinitely many SINR and EH constraints per Rx due to CSI uncertainties, we first reformulate the worst-case SINR and EH constraints into quadratic matrix inequality (QMI) forms. Then, we rely on S-lemma \cite{BoydBook04} and a linear matrix inequality (LMI) representation for robust QMIs \cite{Huang13} to recast the intended robust BFPS problem into a tractable semidefinite program (SDP).

\item To facilitate a {\em decentralized} robust BFPS design, we introduce auxiliary local variables and properly restructuring the formulated SDP. Specifically, leveraging the state-of-the-art ADMM, we develop a decentralized algorithm capable of computing the robust BFPS schemes with local CSI and limited information exchange among the independent Txs. The proposed approach is well suited to the typical MISO SWIPT interference channel setups where there does not exist a central control unit as well as a backhaul link connecting these Txs.

\end{enumerate}

The reminder of this paper is organized as follows. Section~\ref{sec:system} introduces the system model and the robust BFPS design problem under consideration. Section~\ref{sec:robust} reformulates the intended robust BFPS problem. Section~\ref{sec:decentralized} develops the decentralized algorithm for optimal robust BFPS designs relying on the ADMM. Numerical results demonstrate the merit of the proposed scheme in Section~\ref{sec:numerical}, followed by the concluding remarks in Section~\ref{sec:conclusion}.

{\it Notation:} Boldface letters refer to vectors (lower case) or matrices (upper case) and standard lower-case letters denote scalars, unless stated otherwise. $(\cdot)^H$ is the Hermitian operator; $\| \cdot \|$ and $| \cdot |$ stand for the Euclidean norm for a vector and the absolute value of a scalar, respectively; $\mathbb{R}^{x\times y}$ and $\mathbb{C}^{x\times y}$ denote the spaces of real-valued and the complex-valued $x\times y$ matrices, respectively. $\mathbb{E}\{\cdot\}$ is the statistical expectation. ${\rm tr}(\boldsymbol{A})$ and ${\rm rank}(\boldsymbol{A})$ represent the trace of $\boldsymbol{A}$ and the rank of $\boldsymbol{A}$, respectively. A circularly symmetric complex Gaussian random variable $z$ with mean $\mu$ and variance $\sigma^2$ is represented as $\mathcal{CN}(\mu,\sigma^2)$. For a square matrix $\mv A$, $\boldsymbol{A}\succeq \boldsymbol{0}$ indicates that $\boldsymbol{A}$ is positive semidefinite.

\section{System Model and Problem Formulation}\label{sec:system}

\subsection{System Model}

\begin{figure}
\centering
\includegraphics[width=3.5in]{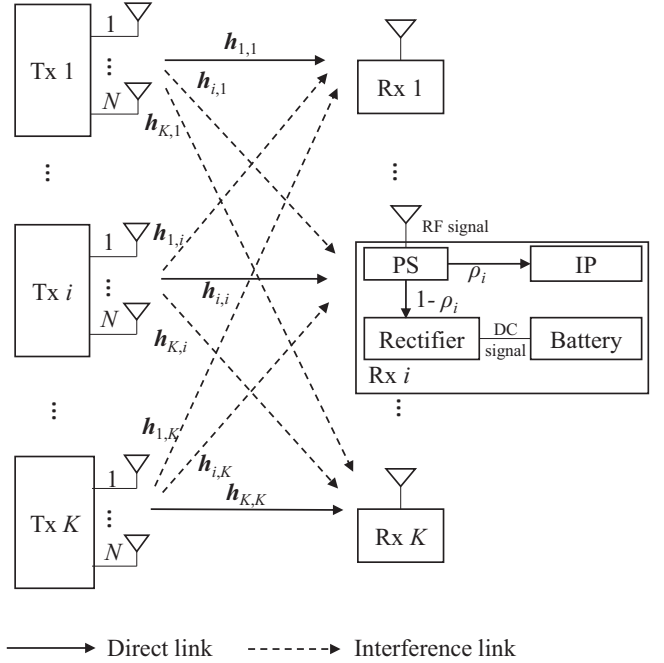}
\caption{System model for $K$-user MISO SWIPT interference channel.}\label{fig:SysMod}
\end{figure}

As shown in Fig. \ref{fig:SysMod}, we consider a $K$-user MISO SWIPT interference channel where the $K$ pairs of Txs and Rxs communicate over the same frequency band. Each Tx $i\in{\cal K}$, equipped with $N>1$ antennas, sends signal to its intended single-antenna Rx-$i$ for both information processing (IP) and EH, where $i\in{\cal K}\triangleq \{1,\ldots,K\}$. Let $\boldsymbol{h}_{i,j}\in\mathbb{C}^{N\times 1}$ denote the $N$-dimensional channel vector from Tx-$j$ to Rx-$i$, $\forall i,j\in{\cal K}$. Then the received baseband signal at Rx-$i$, denoted by $y_i\in\mathbb{C}$, can be expressed as
\begin{align}
 y_i=\boldsymbol{h}_{i,i}^H\boldsymbol{w}_i s_i+\sum_{j=1,\;j\neq i}^K \boldsymbol{h}_{i,j}^H\boldsymbol{w}_js_j+n_i, ~~\forall i\in{\cal K},
\end{align}
where $s_i\in \mathbb{C}$ and $\boldsymbol{w}_i\in\mathbb{C}^{N\times 1}$ denote the transmit signal and transmit beamforming vector intended for Rx-$i$, respectively, and $n_i\sim\mathcal{CN}(0,\sigma_i^2)$ is the additive white Gaussian noise (AWGN) with mean zero and variance $\sigma_i^2$. Without loss of generality, we
assume that $\mathbb{E}\{|s_i|\}=1$, $\forall i\in{\cal K}$, i.e., each transmit signal $s_i$ is of unit power. Hence, the average total transmission power across all $K$ Txs is given by
\begin{equation}
 \sum_{i=1}^K \mathbb{E}\{\|\boldsymbol{w}_is_i\|^2\}=\sum_{i=1}^K \boldsymbol{w}_i^H\boldsymbol{w}_i.
\end{equation}

With a power splitter, each SWIPT Rx-$i$, $\forall i\in{\cal K}$, can split its received signal $y_i$ into two parts, one is used for IP and the other is used for EH (cf. Fig.~\ref{fig:SysMod}). Given a PS ratio $\rho_i\in [0,1]$, the $\rho_i$ portion of the received power is used to decode information and the remaining $(1-\rho_i)$ portion of the received power is used to transfer energy, where $\rho_i\geq 0$, $\forall i\in{\cal K}$. Therefore, the baseband signal for IP at Rx-$i$ can be written as
\begin{equation}
 y_i^{\rm{IP}}  = \sqrt{\rho_i} y_i + v_i,~~\forall i\in{\cal K},
\end{equation}
where $v_i\sim \mathcal{CN}(0,\delta_i^2)$ is the AWGN introduced by the IP circuit. The SINR for Rx-$i$ is thus given by
\begin{align}\label{eq.sinr}
 &{\rm SINR}_i\left(\{\mv h_{i,j},\mv w_j\},\rho_i\right) \notag \\
 &~~~~~~~\triangleq \frac{|\boldsymbol{h}_{i,i}^H\boldsymbol{w}_i|^2}{\sum_{j=1,j\neq i}^K |\boldsymbol{h}_{i,j}^H\boldsymbol{w}_j|^2+\sigma_i^2+\delta_i^2/\rho_i}, ~~\forall i\in{\cal K}.
\end{align}

On the other hand, the equivalent baseband signal split into the EH circuit for Rx-$i$ is
\begin{equation}
 y_i^{\rm{EH}} = \sqrt{1-\rho_i}y_i ,~~\forall i\in{\cal K}.
\end{equation}
Note that the RF signal received by each Rx-$i$, $\forall i \in{\cal K}$, is first converted to a direct current (DC) signal by a rectifier and then the energy of the DC signal is then stored in the chargeable battery (cf.~Fig.~\ref{fig:SysMod}). During this EH process, some amount of energy is lost mainly due to the imperfection of the rectifier. Assuming a linear EH model,\footnote{The recently studied non-liner EH model and waveform design for efficient WPT (see, e.g., \cite{Zeng17} and references therein) are beyond the scope of this paper.} i.e., with a constant RF-to-DC energy conversion efficiency, the energy harvested by the EH circuit at Rx-$i$ is \cite{Zhang13}
\begin{align}\label{eq.eh}
 {\rm EH}_i(\{\mv h_{i,j},\mv w_i\},\rho_i) &\triangleq \zeta_i\mathbb{E}\{|y_i^{\rm{EH}}|^2\} \notag\\
 &=\zeta_i(1-\rho_i)\bigg(\sum_{j=1}^K |\boldsymbol{h}_{i,j}^H\boldsymbol{w}_j|^2 +\sigma_i^2\bigg) \notag \\
 & \quad \quad \quad \quad \quad \quad \quad \quad \quad \quad \quad \forall i\in{\cal K},
\end{align}
where ${\zeta}_i\in (0,1]$ denotes the constant EH efficiency per Rx-$i$ corresponding to the ratio of the harvested energy to the received energy.

\subsection{Channel Uncertainty}
As with the existing works in robust beamforming designs\cite{Xiang12,Zhao15,Feng15,Ng14}, we consider an additive CSI uncertainty model, where the channel is modeled as a random vector variable. This model is widely employed to account for the CSI uncertainty at the transmitter due to estimation errors, feedback quantization, and delays, etc.
Let $\hat{\boldsymbol{h}}_{i,j}\in\mathbb{C}^{N\times 1}$, $\forall i,j\in{\cal K}$, be the estimated channels at the $K$ Txs. Each Tx $i$ treats this ``nominal channel'' $\hat{\boldsymbol{h}}_{i,j}$, $\forall j\in{\cal K}$, as deterministic, and it adds ``perturbation'' terms in order to account for CSI uncertainty. Then the imperfect CSI of the true channels can be perceived as
\begin{equation}\label{eq.csi}
\boldsymbol{h}_{i,j}=\hat{\boldsymbol{h}}_{i,j}+\boldsymbol{e}_{i,j}, ~~\forall i, j\in{\cal K},
\end{equation}
where $\boldsymbol{e}_{i,j}\in{\mathbb C}^{N\times 1}$ is a random vector variable. In this paper, we assume that each term $\boldsymbol{e}_{i,j}\in{\mathbb C}^{N\times 1}$ is bounded by an ellipsoid $\mathcal{E}_{i,j}$:
\begin{equation}
\mathcal{E}_{i,j}\triangleq \left\{\boldsymbol{e}_{i,j} |~\boldsymbol{e}_{i,j}^H\boldsymbol{B}_{i,j}\boldsymbol{e}_{i,j} \leq 1 \right\},~~\forall i, j\in{\cal K},
\end{equation}
where $\boldsymbol{B}_{i,j}\succeq \boldsymbol{0}$ determines the size and shape of the error ellipsoid and is assumed to be available to the Txs.

\subsection{Problem Formulation}
Our objective in this paper is to optimize the transmit beamforming and receive PS ratio per user-pair in Fig.~\ref{fig:SysMod} based on imperfecrt CSI available at the $K$ Txs. Specifically, we aim to minimize the total transmission power subject to both the worst-case SINR and EH constraints under the CSI uncertainty model \eqref{eq.csi}. Let $\gamma_i$ and $\eta_i$ stand for the prescribed SINR and EH targets for each Rx $i\in{\cal K}$, respectively. The robust BFPS design for jointly finding the optimal beamforming vectors $\mv w_i$'s and PS ratios $\rho_i$'s is then formulated as:
\begin{subequations}\label{eq.wc}
\begin{align}
 ({\rm P}1):~
 \min_{\{\boldsymbol{w}_i,\rho_i\}} &~ \sum_{i=1}^K \boldsymbol{w}_i^H\boldsymbol{w}_i\\
~~ ~~\text{s.t.}~~&
{\rm SINR}_i(\{\hat{\mv h}_{i,j}+\mv e_{i,j}, \mv w_i\},\rho_i) \geq \gamma_i \notag\\
& \quad\quad\quad ~~\forall \mv e^H_{i,j}\mv B_{i,j}\mv e_{i,j}\leq 1,~~\forall i,j\in{\cal K} \\
&~{\rm EH}_i(\{\hat{\mv h}_{i,j}+\mv e_{i,j}, \mv w_i\},\rho_i) \geq \eta_i \notag \\
& \quad\quad\quad~~\mv e^H_{i,j}\mv B_{i,j}\mv e_{i,j}\leq 1,~~\forall i,j\in{\cal K}  \\
& ~0\leq \rho_i\leq 1, ~~\forall i\in{\cal K},
\end{align}
\end{subequations}
where the constraints in (\ref{eq.wc}b) and (\ref{eq.wc}c), a.k.a. the {\em worst-case} SINR and EH constraints, guarantee information transmission reliability and EH reliability per Rx-$i$, respectively, for arbitrary channel realizations $\{\mv h_{i,j}\}$ by \eqref{eq.csi}; this is a conservative but safe design fashion to counter against CSI uncertainties. Note that the number of the SINR and EH constraints in (P1) is infinite due to the randomness and continuous of $\mv e_{i,j}$. Problem (P1) is a semi-infinite optimization problem and very difficult to solve\cite{BoydBook04}. To make (P1) tractable, we first transform these infinite-many constraints into equivalent QMIs, and then establish the LMI relaxation (or termed as SDP relaxation) problem in the next section.

\section{Problem Reformulation for (P1)}\label{sec:robust}
In this section, we reformulate the robust BFPS design problem (P1) into a tractable SDP based on S-lemma and equivalent LMI representation for QMI. 

To this end, the worst-case SINR constraint per Rx-$i$ in (\ref{eq.wc}) is first given as
\begin{align}\label{eq.wc_sinr_v1}
& \frac{|(\boldsymbol{\hat h}_{i,i}+e_{i.i})^H\boldsymbol{w}_i|^2}{\sum_{j=1,j\neq i}^K |(\boldsymbol{\hat h}_{i,j}+\mv e_{i,j})^H\boldsymbol{w}_j|^2+\sigma_i^2+\delta_i^2/\rho_i}\geq \gamma_i, \notag \\
& \quad\quad\quad\quad\quad\quad  \quad\quad\quad\quad ~~\forall \mv e^H_{i,j}\mv B_{i,j}\mv e_{i,j}\leq 1,~~\forall i\in{\cal K}.
\end{align}
With simple algebraic manipulation, \eqref{eq.wc_sinr_v1} can be rewritten as
\begin{align}\label{eq.wc_sinr_v2}
&\frac{|(\boldsymbol{\hat h}_{i,i}+e_{i.i})^H\boldsymbol{w}_i|^2}{\gamma_i}-\sum_{j=1,j\neq i}^K |(\boldsymbol{\hat h}_{i,j}+\mv e_{i,j})^H\boldsymbol{w}_j|^2-\sigma_i^2 \geq \frac{\delta_i^2}{\rho_i},
\notag \\
& \quad\quad\quad\quad\quad\quad \quad \quad \quad\quad ~~\forall \mv e^H_{i,j}\mv B_{i,j}\mv e_{i,j}\leq 1,~~\forall i\in{\cal K}.
\end{align}
Define $\boldsymbol{W}_i \triangleq \boldsymbol{w}_i\boldsymbol{w}_i^H$, $\forall i\in{\cal K}$. Using the Schur-complement condition for positive semi-definiteness \cite{BoydBook04}, the worst-case SINR constraints in \eqref{eq.wc_sinr_v2} can be expressed as the following QMIs (with respect to (w.r.t.) $\{\mv e_{i,j}\}$):
\begin{align}\label{eq.rho_i}
&  \begin{bmatrix}
 \rho_i& \delta_i\\
 \delta_i & \frac{(\hat{\boldsymbol{h}}_{i,i}+\boldsymbol{e}_{i,i})^H\boldsymbol{W}_i(\hat{\boldsymbol{h}}_{i,i}+\boldsymbol{e}_{i,i})}{\gamma_i}-\sum_{j=1,j\neq i}^K \bar{t}_{i,j}-\sigma_i^2
\end{bmatrix}
 \succeq \boldsymbol{0},\notag \\
 &\quad\quad\quad\quad\quad\quad\quad \quad\quad\quad \forall \mv e^H_{i,j}\mv B_{i,j}\mv e_{i,j}\leq 1,~~ \forall i\in{\cal K},
 \end{align}
where $\bar{t}_{i,j}\geq 0$, $\forall i,j\in{\cal K}$, are the introduced slack variables and satisfy the infinitely many QMIs (w.r.t. $\{\mv e_{i,j}\}$):
\begin{align}\label{eq.e_ij}
 &(\boldsymbol{\hat{h}}_{i,j}+\boldsymbol{e}_{i,j})^H\boldsymbol{W}_j(\boldsymbol{\hat{h}}_{i,j}+\boldsymbol{e}_{i,j})\leq \bar{t}_{i,j},\notag \\
 &\quad\quad\quad\quad\quad\quad\quad\quad~~ \forall \mv e^H_{i,j}\mv B_{i,j}\mv e_{i,j}\leq 1,~~\forall i,j\in{\cal K}.
\end{align}
We notice that the number of constraints in \eqref{eq.rho_i} is still infinite. Fortunately, resorting to \cite[Corollary 4.3]{Huang13}, we can obtain the equivalent LMI representation for the QMIs \eqref{eq.rho_i}. For completeness, we borrow the following lemma here:

\begin {lemma}(Corollary 4.3, \cite{Huang13})\label{lem:lem1}
If $\boldsymbol{D}_i\succeq \boldsymbol{0}$, $i\in\{1,2\}$, then the following QMI system
\begin{align}\label{eq.QMI}
&\begin{bmatrix}
\boldsymbol{A}_1 & \boldsymbol{A}_2+\boldsymbol{A}_3\boldsymbol{X}\\
(\boldsymbol{A}_2+\boldsymbol{A}_3\boldsymbol{X})^H & \boldsymbol{A}_4+\boldsymbol{A}_5\boldsymbol{X}+(\boldsymbol{A}_5\boldsymbol{X})^H+\boldsymbol{X}^H\boldsymbol{A}_6\boldsymbol{X}
\end{bmatrix} \succeq\boldsymbol{0},
\notag \\
&~~~~~~~~~~~~~~~~~~~~~~~ \boldsymbol{X}: {\rm tr}(\boldsymbol{D}_i\boldsymbol{X}\boldsymbol{X}^H)\leq 1,~ i=1,2
\end{align}
is equivalent to the following LMI system: there exist $\nu_1\geq 0$ and $\nu_2\geq 0$, such that
\begin{align}\label{eq.LMI}
\begin{bmatrix}
\boldsymbol{A}_1&\boldsymbol{A}_2&\boldsymbol{A}_3 \\
\boldsymbol{A}_2^H&\boldsymbol{A}_4&\boldsymbol{A}_5\\
\boldsymbol{A}_3^H&\boldsymbol{A}_5^H&\boldsymbol{A}_6
\end{bmatrix}
-\nu_1\begin{bmatrix}
\boldsymbol{0}&\boldsymbol{0}&\boldsymbol{0}\\
\boldsymbol{0}&\boldsymbol{I}&\boldsymbol{0}\\
\boldsymbol{0}&\boldsymbol{0}&-\boldsymbol{D}_1
\end{bmatrix}\notag \\
-\nu_2\begin{bmatrix}
\boldsymbol{0}&\boldsymbol{0}&\boldsymbol{0}\\
\boldsymbol{0}&\boldsymbol{I}&\boldsymbol{0}\\
\boldsymbol{0}&\boldsymbol{0}&-\boldsymbol{D}_2
\end{bmatrix}
\succeq \boldsymbol{0},
\end{align}
where the parameters $\mv A_i$ and $\mv D_i$ could be of any proper dimensions so that the related system are well-defined.
\end{lemma}

Exploiting Lemma~\ref{lem:lem1} and setting $\mv X=\mv e_{i,i}$, $\mv A_1=\rho_i$, $\mv A_2=\delta_i$, $\mv A_3=\mv 0$, $\mv A_4=\frac{\hat{\boldsymbol{h}}_{i,i}^H{\boldsymbol{W}_i}\hat{\boldsymbol{h}}_{i,i}}{\gamma_i}-\sum_{j=1,j\neq i}^K \bar{t}_{i,j}-\sigma_i^2$, $\mv A_5=\frac{\hat{\boldsymbol{h}}_{i,i}^H\boldsymbol{W}_i}{\gamma_i}$, $\mv A_6=\frac{\mv W_i}{\gamma_i}$, $\mv D_1=\mv B_{i,i}$, and $\mv D_2=\mv 0$, $\forall i\in{\cal K}$, it follows that the equivalent LMI representations for \eqref{eq.rho_i} are: there exist $ \nu_{i}\geq 0$, $\forall i\in{\cal K}$, such that
\begin{align}\label{eq.rho_i1}
&{\begin{bmatrix}
 \rho_i & \delta_i &\boldsymbol{0} \\
 \delta_i &\frac{\hat{\boldsymbol{h}}_{i,i}^H{\boldsymbol{W}_i}\hat{\boldsymbol{h}}_{i,i}}{\gamma_i}-\sum_{j\neq i} \bar{t}_{i,j}-\sigma_i^2  -\nu_{i}& \frac{\hat{\boldsymbol{h}}_{i,i}^H\boldsymbol{W}_i}{\gamma_i}\\
 \boldsymbol{0} &\frac{\boldsymbol{W}_i\hat{\boldsymbol{h}}_{i,i}}{\gamma_i}& \frac{\boldsymbol{W}_i}{\gamma_i}+\nu_{i}\boldsymbol{B}_{i,i}
\end{bmatrix}} \succeq \mv 0, \notag \\
&~~~~~~~~~~~~~~~~~~~~~~~~~~~~~~~~~~~~~~~~~~~~~~~~~~~~~\forall i\in{\cal K}.
 \end{align}
Note that the number of constraints in \eqref{eq.e_ij} is infinite and the S-lemma can be applied for reformulation, which is introduced as the following lemma:

\begin{lemma}(S-lemma,\cite{BoydBook04})\label{lem:S-lem}
Let $\mv A$ and $\mv B$ be two $n\times n$ Hermitian matrices, $\mv c\in {\mathbb C}^{n\times 1}$, and $d\in{\mathbb R}$. Then the following two conditions are equivalent:
\begin{enumerate}
 \item $\mv x^H \mv A \mv x+\mv c^H \mv x+ \mv x^H \mv c+ d\geq 0$ for all $\mv x\in{\mathbb C}^{n\times1}$ such that $\mv x^H \mv B\mv x \leq 1$;
\item There exists a $\lambda \in {\mathbb R}$ such that
\begin{align}
\lambda \geq 0,\quad
\begin{bmatrix}
\mv A+\lambda\mv B & \mv c\\
\mv c^H & d-\lambda
\end{bmatrix} \succeq \mv 0.
\end{align}
\end{enumerate}
\end{lemma}
Based on Lemma~\ref{lem:S-lem}, by setting $\mv X=\mv e_{i,j}$, $\mv A = -\mv W_j$, $\mv c= -\mv W_j \hat{\mv h}_{i,j}$, $d=\bar{t}_{i,j}-\hat{\mv h}^H_{i,j}\mv W_j \hat{\mv h}_{i,j}$, and $\mv B=\mv B_{i,j}$, $\forall i,j\in{\cal K}$, the inequalities in \eqref{eq.e_ij} can be recast into the LMIs: there exists a $\lambda_{i,j} \ge 0$ such that
\begin{align}\label{eq.rho_i2}
&\begin{bmatrix}
 -\boldsymbol{W}_j+{\lambda_{i,j}}\boldsymbol{B}_{i,j} & -\boldsymbol{W}_j\boldsymbol{\hat{h}}_{i,j}\\
 -\hat{\boldsymbol{h}}^H_{i,j}\boldsymbol{W}_j^H & \bar{t}_{i,j}-\hat{\boldsymbol{h}}_{i,j}^H\boldsymbol{W}_j\hat{\boldsymbol{h}}_{i,j}-\lambda_{i,j}
\end{bmatrix} \succeq \boldsymbol{0}, ~\forall i,j\in{\cal K}.
 \end{align}

Regarding the worst-case EH constraints in (\ref{eq.wc}b), the explicit expressions are given as
\begin{align}\label{eq.wc_EH_v1}
&\zeta_i(1-\rho_i)\bigg(\sum_{j=1}^K |(\boldsymbol{h}_{i,j}+\mv e_{i,j})^H\boldsymbol{w}_j|^2 +\sigma_i^2\bigg)\geq \eta_i,\notag \\
&\quad\quad\quad\quad\quad\quad\quad\quad\quad \forall \mv e^H_{i,j}\mv B_{i,j}\mv e_{i,j}\leq 1,~~\forall i,j\in{\cal K}.
\end{align}
By the Schur-complement condition for positive semi-definiteness \cite{BoydBook04}, \eqref{eq.wc_EH_v1} can be rewritten as the following LMIs:
\begin{align}\label{eq.eh1}
&{\begin{bmatrix}
 \zeta_i(1-\rho_i) & \sqrt{\eta_i}\\
 \sqrt{\eta_i} & \sum_{j=1}^K\underline{t}_{i,j}+\sigma_i^2
\end{bmatrix}} \succeq \boldsymbol{0}, ~~\forall i\in{\cal K},
\end{align}
where $\underline{t}_{i,j}\geq 0$, $\forall i,j\in{\cal K}$, are the introduced slack variables and satisfy the infinite-many QMIs (w.r.t. $\{\mv e_{i,j}\}$):
\begin{align}\label{eq.min2}
&(\hat{\boldsymbol{h}}_{i,j}+\boldsymbol{e}_{i,j})^H\boldsymbol{W}_j(\hat{\boldsymbol{h}}_{i,j}+\boldsymbol{e}_{i,j})\geq \underline{t}_{i,j},\notag \\
&\quad\quad\quad\quad\quad\quad\quad\quad~~\forall \mv e^H_{i,j}\mv B_{i,j}\mv e_{i,j}\leq 1,~~\forall i,j\in{\cal K}.
\end{align}
Again by Lemma~\ref{lem:S-lem} and setting $\mv X=\mv e_{i,j}$, $\mv A = \mv W_j$, $\mv c= \mv W_j \hat{\mv h}_{i,j}$, $d=\hat{\mv h}^H_{i,j}\mv W_j \hat{\mv h}_{i,j}-\underline{t}_{i,j}$, and $\mv B=\mv B_{i,j}$, $\forall i,j\in{\cal K}$, we can recast \eqref{eq.min2} into the LMIs: there exists an $\mu_{i,j} \ge 0$ such that
\begin{align}\label{eq.eh1_1}
 &\begin{bmatrix}
 \boldsymbol{W}_j+{\mu_{i,j}}\boldsymbol{B}_{i,j} & \boldsymbol{W}_j\boldsymbol{\hat{h}}_{i,j}\\
\hat{\boldsymbol{h}}_{i,j}^H\boldsymbol{W}_j^H & \hat{\boldsymbol{h}}_{i,j}^H\boldsymbol{W}_j\hat{\boldsymbol{h}}_{i,j}-\underline{t}_{i,j}-\mu_{i,j}
 \end{bmatrix}  \succeq \boldsymbol{0}, ~~\forall i,j\in{\cal K}.
 \end{align}

With the obtained LMIs \eqref{eq.rho_i1}, \eqref{eq.rho_i2}, \eqref{eq.eh1}, and \eqref{eq.eh1_1}, we are ready to reformulate (P1) into a rank-constrained SDP:
\begin{align*}\label{eq.wc2}
({\rm P2}):& \min_{\{\boldsymbol{W}_i,\rho_i,\nu_i,\lambda_{i,j},\mu_{i,j},\bar{t}_{i,j},\underline{t}_{i,j}\}} ~~{\sum_{i=1}^{K}{\rm tr}(\boldsymbol{W}_i)}\\
&\text{s.t. } \eqref{eq.rho_i1}, ~\eqref{eq.rho_i2},~\eqref{eq.eh1}, ~{\rm and}~\eqref{eq.eh1_1} \notag \\
&~~~~\lambda_{i,j}\geq0,~ \mu_{i,j}\geq 0,~\bar{t}_{i,j} \geq 0,~ \underline{t}_{i,j}\geq 0, ~~\forall i, j\in{\cal K}\notag \\
&~~~~\nu_i \geq 0, ~0\leq \rho_i \leq1,~~\forall i\in{\cal K} \notag \\
&~~~~\boldsymbol{W}_i \succeq \boldsymbol{0}, ~{\rm rank}(\boldsymbol{W}_i)=1,~~\forall i\in{\cal K}.\notag
\end{align*}
Note that problem (P2) is still non-convex only due to the rank-one constraints. By dropping them, (P2) becomes an SDP and can thus be efficiently solved by off-the-shelf convex solvers such as CVX\cite{CVX}. Denote by $\boldsymbol{W}_i^*$, $\forall i\in{\cal K}$, the optimal matrices for (P2). If ${\rm rank}(\boldsymbol{W}_i^*)=1$, $\forall i\in{\cal K}$, the robust optimal beamformers $\boldsymbol{w}_i^*$, $\forall i\in{\cal K}$, can be exactly obtained by eigenvalue
decomposition (EVD), i.e., $\boldsymbol{W}_i^*=\boldsymbol{w}_i^*\boldsymbol{w}_i^{*H}$. Otherwise, the relaxed problem only provides a lower bound for the minimum transmission power, and a Gaussian randomization procedure can be applied to find an approximate solution \cite{Gau}.

\begin{remark}
Note that a recent work \cite{Zhao15} considered a similar robust BFPS design and independently developed another SDP relaxation for (P1) (see eqn. (21) in \cite{Zhao15}). Specifically, by replacing $1/\rho_i$ and $1/(1-\rho_i)$ with auxiliary variables $a_i\geq 0$ and $b_i\geq 0$, and adding convex constraints $1/a_i+1/b_i \leq 1$, $\forall i\in{\cal K}$, the worst-case SINR and EH constraints can be transformed into LMIs via the S-lemma, leading to another SDP relaxation. Compared to our relaxation problem, (P2) without the rank-one constraints, the difference lies in that in our reformulation, $\rho_i$ in every constraint is incorporated into a corresponding LMI constraint (as in \eqref{eq.rho_i} and \eqref{eq.eh1}) via Schur-complement Lemma. More importantly, beyond the convex optimization methods (as in \cite{Zhao15}), we herein establish an approach to obtain the decentralized solutions to the matrix form (P2) of the original problem (P1), as we pursue next.
\end{remark}

\section{Decentralized ADMM Approach for (P2)}\label{sec:decentralized}

As discussed in the previous section, directly solving problem (P2) requires a central controller with global CSI estimates at the Txs. However, for typical interference channel setups, there may not exist a central control unit as well as a backhaul link connecting the multiple distributed Txs. Hence, a decentralized solution for (P2) is clearly a need. For this reason, we next leverage the celebrated ADMM \cite{Boyd08} to pursue an optimal robust BFPS design scheme in a decentralized fashion. The more details of ADMM are referred to \cite{Boyd08}. In what follows, we first introduce ADMM and then present the ADMM based decentralized BFPS algorithm, followed by its convergence analysis.

\subsection{Introduction of ADMM}
  ADMM blends the decomposability of dual ascent with the superior convergence properties of the method of multiplier; it takes the form of a {\em decomposition-coordination} procedure, where the solutions of the small local subproblems are coordinated to find the solution of the large global problem. Consider the following convex problem \cite{Boyd08}
  \begin{subequations}\label{eq.admm1}
  \begin{align}
  \min_{\mv x,\mv z} ~& ~f(\mv x)+g(\mv z)\\
  {\rm s.t.}~~& \mv A\mv x+\mv B\mv z = \mv d,
  \end{align}
  \end{subequations}
  where $f:\mathbb{R}^{n\times 1}\longmapsto \mathbb{R}$ and $g:\mathbb{R}^{m\times 1} \longmapsto \mathbb{R}$ are convex functions; $\mv A\in\mathbb{R}^{p\times n}$, $\mv B\in\mathbb{R}^{p\times m}$, and $\mv d \in\mathbb{R}^{p\times 1}$.

By introducing the Lagrangian multipliers $\mv y \in \mathbb{R}^{p\times 1} $ for (\ref{eq.admm1}b) and a penalty parameter $c>0$, ADMM considers the following augmented Lagrangian:
 \begin{align}\label{eq.admm2}
    L_c(\mv x,\mv z,\mv y) = & f(\mv x)+ g(\mv z) +\mv y^H(\mv A \mv x+\mv B\mv z-\mv c)\notag\\
    &+\frac{c}{2}\|\mv A \mv x+\mv B\mv z-\mv d\|^2,
 \end{align}
 and consists of the iterations:
 \begin{subequations}\label{eq.admm3}
 \begin{align}
 \mv x^{q+1} &= \argmin_{\mv x} L_c(\mv x,\mv z^{q},\mv y^{q})\\
 \mv z^{q+1} &= \argmin_{\mv x} L_c(\mv x^{q+1},\mv z^{q},\mv y^{q})\\
 \mv y^{q+1} &= \mv y^q + c(\mv A \mv x^{q+1}+\mv B\mv z^{q+1}-\mv d) ,
 \end{align}
 \end{subequations}
 where $q\geq 0$ denotes the iteration index. As in \eqref{eq.admm3}, $\mv x$ and $\mv z$ are updated in an alternating sequential fashion accounting for {\em alternating direction} in ADMM.

 The ADMM can convergence to the global solution of \eqref{eq.admm1} under the following conditions (see Section 3.2 in \cite{Boyd08}):
 \begin{enumerate}
 \item The epigraphs of the functions $f$ and $g$ are closed and nonempty convex sets.
  \item Strong duality holds for \eqref{eq.admm1} and its Lagrangian dual problem.
  \end{enumerate}

\subsection{Decentralized Approach for (P2) based on ADMM}
In this subsection, we present the proposed decentralized approach for (P2) via ADMM. To this end, we first introduce the following slack variables:
\begin{equation}
p_i \triangleq {\rm tr}(\boldsymbol{W}_i), ~~\overline{T}_{i} \triangleq \sum_{j=1,j\neq i}^K\bar{t}_{i,j}, ~~\underline{T}_i \triangleq \sum_{j=1,j\neq i}^K \underline{t}_{i,j},~~\forall i\in{\cal K},
\end{equation}
where $p_i$ is interprated as the transmission power of Tx-$i$; $\overline{T}_{i}$ and $\underline{T}_i$ represent the sum of the required maximum and minimum interference from the remaining $(K-1)$ Txs to Rx-$i$, respectively. Note that we can equivalently interchange the subindices $i$ and $j$ in \eqref{eq.rho_i2} and \eqref{eq.eh1_1} due to the fact of $i,j\in{\cal K}$.

Then, by dropping the rank-one constraints, the constraints in (P2) can be decomposed into $K$ independent convex sets:
\begin{align}
&\mathcal{C}_i \triangleq \notag \\
&\Bigg\{\left(\mv\Psi_i,\{\bar{t}_{j,i},\underline{t}_{j,i}\}_j,\overline{T}_i,\underline{T}_i \right)~\Big| \notag\\
&~\begin{bmatrix}
 -\boldsymbol{W}_i+\lambda_{j,i}\boldsymbol{B}_{j,i} & -\boldsymbol{W}_i\hat{\boldsymbol{h}}_{j,i}\\
 -\hat{\boldsymbol{h}}_{j,i}^H\boldsymbol{W}_i^H & \bar{t}_{j,i}-\hat{\boldsymbol{h}}_{j,i}^H\boldsymbol{W}_i\hat{\boldsymbol{h}}_{j,i}-\lambda_{j,i}
\end{bmatrix}
 \succeq \boldsymbol{0},~\forall j\in{\cal K}\setminus i, \notag \\
&~\begin{bmatrix}
 \boldsymbol{W}_i+\mu_{j,i}\boldsymbol{B}_{j,i} & \boldsymbol{W}_i\hat{\boldsymbol{h}}_{j,i}\\
\hat{\boldsymbol{h}}_{j,i}^H\boldsymbol{W}_i^H & \hat{\boldsymbol{h}}_{j,i}^H\boldsymbol{W}_i\hat{\boldsymbol{h}}_{j,i}-\underline{t}_{j,i}-\mu_{j,i} \end{bmatrix}  \succeq \boldsymbol{0},~ \forall j\in {\cal K}\setminus i,\notag \\
&~~\lambda_{j,i} \geq 0,~\mu_{j,i}\geq 0,~ \bar{t}_{j,i} \geq 0, ~\underline{t}_{j,i}\geq 0, ~~\forall j\in{\cal K},\notag \\
&~~ \overline{T}_{i}\geq 0,~ \underline{T}_i\geq 0, ~0\leq \rho_i \leq1,~ \nu_i\geq 0,~\boldsymbol{W}_i \succeq \boldsymbol{0}, \notag\\
&~~p_i= {{\rm tr}(\boldsymbol{W}_i)},~ \eqref{eq.rho_i1}, ~ {\rm{and}}~ \eqref{eq.eh1} \notag \Bigg\},~~\forall i\in{\cal K},
\end{align}
where $\mv \Psi_i\triangleq \{\boldsymbol{W}_i,\rho_i,\nu_i,\{\lambda_{j,i}\}_{j=1}^K, \{\mu_{j,i}\}_{j=1}^K\}$, $\forall i\in{\cal K}$. Furthermore, by excluding the terms $\bar{t}_{i,i}$ and $\underline{t}_{i,i}$, the local vector $\boldsymbol{t}_i\in\mathbb{R}^{2K\times 1}$ maintained by Tx-$i$ is defined as
\begin{align}\label{eq.local}
\boldsymbol{t}_i \triangleq \Big[~ &\overline{T}_i,\underbrace{\bar{t}_{1,i},\ldots,\bar{t}_{i-1,i},\bar{t}_{i+1,i},\ldots,\bar{t}_{K,i}}_{(K-1)~\text{terms}},\notag \\
&\underline{T}_i,\underbrace{\underline{t}_{1,i},\ldots,\underline{t}_{i-1,i},\underline{t}_{i+1,i},\ldots,\underline{t}_{K,i}}_{(K-1) ~\text{terms}}\Big]^H, ~~\forall i\in{\cal K},
\end{align}
and let a ``public'' vector $\boldsymbol{t}\in\mathbb{R}^{2K(K-1)\times 1}$  collect all $\bar{t}_{i,j}$ and $\underline{t}_{i,j}$:
\begin{align}\label{eq.public}
\boldsymbol{t} \triangleq \Big[~&\underbrace{\bar{t}_{1,2},\ldots,\bar{t}_{1,K},\ldots,\bar{t}_{K,1},\ldots,\bar{t}_{K,K-1}}_{K(K-1)~\text{terms}},\notag \\
&\underbrace{\underline{t}_{1,2},\ldots,\underline{t}_{1,K},\ldots,\underline{t}_{K,1},\ldots,\underline{t}_{K,K-1}}_{K(K-1)~\text{terms}}\Big]^H.
\end{align}

 Based on \eqref{eq.local} and \eqref{eq.public}, there exists one linear mapping matrix $\boldsymbol{E}_i \in\{0,1\}^{2K\times 2K(K-1)}$ such that
$\boldsymbol{t}_i=\boldsymbol{E}_i\boldsymbol{t}$, $\forall i\in {\cal K}$.
Then, the SDP relaxation of (P2) can be rewritten in a compact form:
\begin{subequations}\label{eq.wc4}
\begin{align}
\min_{\{\mv \Psi_i,\boldsymbol{t}_{i},\boldsymbol{t} \} }
 &~\sum_{i=1}^{K}p_i  \\
 \text{s.t. }&
(\mv \Psi_i,\boldsymbol{t}_{i} ) \in\mathcal{C}_i,~ \boldsymbol{t}_i=\boldsymbol{E}_i\boldsymbol{t},~~\forall i\in{\cal K}.
\end{align}
\end{subequations}
Note that, by incorporating the rank-one constraints of $\rm{rank}(\mv W_i)=1$, $\forall i\in{\cal K}$, problem \eqref{eq.wc4} is equivalent to (P2). In the following, we employ the ADMM method (cf. Section IV.A) for solving \eqref{eq.wc4}. Consider the following problem:
\begin{subequations}\label{eq.wc5}
\begin{align}
\min_{\{\mv \Psi_i,\boldsymbol{t}_{i},\boldsymbol{t}\}} & ~\sum_{i=1}^{K}p_i+\frac{c}{2}\sum_{i=1}^{K}||\boldsymbol{E}_i\boldsymbol{t}-\boldsymbol{t}_i||^2\\
\text{s.t. }
& \left(\mv \Psi_i,\boldsymbol{t}_{i} \right) \in\mathcal{C}_i,~\boldsymbol{t}_i=\boldsymbol{E}_i\boldsymbol{t},~~\forall i\in{\cal K},
\end{align}
\end{subequations}
where $c>0$ is again a constant denoting a penalty for the term $\sum_{i=1}^{K}||\boldsymbol{E}_i\boldsymbol{t}-\boldsymbol{t}_i||^2$. Clearly problem (\ref{eq.wc5}) is equivalent to problem (\ref{eq.wc4}), since for any feasible $\{\mv t,\mv t_1,\ldots,\mv t_K\}$ it always holds that $\sum_{i=1}^{K}||\boldsymbol{E}_i\boldsymbol{t}-\boldsymbol{t}_i||^2=0$. The (redundant) penalty terms $(c/2)\sum_{i=1}^{K}||\boldsymbol{E}_i\boldsymbol{t}-\boldsymbol{t}_i||^2$ is introduced to make the associated dual function of \eqref{eq.wc5} can be shown to be differentiable under rather mild conditions on the original problem \eqref{eq.wc4}, thereby facilitating the development of ADMM \cite{Boyd08,BerBook89}. Introducing the Lagrange dual variables $\boldsymbol{\beta}_i\in\mathbb{R}^{2K\times 1}$ associated with the constraints $\boldsymbol{t}_i=\boldsymbol{E}_i\boldsymbol{t}$, $\forall i\in{\cal K}$, the Lagrangian dual problem of (\ref{eq.wc5}) (i.e., the augmented Lagrangian dual problem of \eqref{eq.wc4}) is given by \cite{BoydBook04}
\begin{subequations}\label{eq.dual_prob}
\begin{align}
&\max_{\{\boldsymbol{\nu}_i\}}\min_{\{\mv \Psi_i,\boldsymbol{t}_{i},\boldsymbol{t}\}} \sum_{i=1}^{K} \left(p_i+\frac{c}{2}||\boldsymbol{E_i}\boldsymbol{t}-\boldsymbol{t}_i||^2+\boldsymbol{\beta}_i^H(\boldsymbol{E}_i\boldsymbol{t}-\boldsymbol{t}_i)\right) \\
&~~~~~~~~~~\text{s.t. } \left(\mv \Psi_i,\boldsymbol{t}_i\right)\in\mathcal{C}_i,~~\forall i\in{\cal K}.
\end{align}
\end{subequations}
For problem \eqref{eq.dual_prob}, we can then rely on the ADMM algorithm to develop a decentralized solution as follows.

\underline{ADMM}: For all Tx $i\in{\cal K}$,
initialize dual variables $\{\boldsymbol{\beta}_i(0)\}$ and inner variables $\{\boldsymbol{t}(0), \boldsymbol{t}_i(0)\}$; choose a penalty parameter $c>0$; set the iteration index $q=0$;

At every iteration $q$, do:
\begin{itemize}
\item \textbf{Updating inner variables $\boldsymbol{t}_i$}:  Fix the variable $\boldsymbol{t}(q)$ and
solve the following problem to obtain $\{\boldsymbol{t}_i(q+1)\}$:
\begin{subequations}\label{eq.min_0}
\begin{align}
\min_{\{\mv \Psi_i,\boldsymbol{t}_i\}}&~
\sum_{i=1}^K \left(p_i+\frac{c}{2}||\boldsymbol{E_i}\boldsymbol{t}(q)-\boldsymbol{t}_i||^2 - \boldsymbol{\beta}_i^H(q)\boldsymbol{t}_i\right)\\
\text{s.t. } &
\left(\mv \Psi_i,\boldsymbol{t}_{i} \right) \in\mathcal{C}_i,~~\forall i\in{\cal K}.
\end{align}
\end{subequations}
Note that \eqref{eq.min_0} can be decomposed into $K$ convex subproblems, each given by
\begin{subequations}\label{eq.LocalBeam}
\begin{align}
\min_{\mv \Psi_i,\boldsymbol{t}_i} &~ p_i+\frac{c}{2}||\boldsymbol{E_i}\boldsymbol{t}(q)-\boldsymbol{t}_i||^2 - \boldsymbol{\beta}_i^H(q)\boldsymbol{t}_i\\
\text{s.t. }&
\left(\mv \Psi_i,\boldsymbol{t}_{i} \right) \in\mathcal{C}_i,
\end{align}
\end{subequations}
where $i\in{\cal K}$. Essentially, given local CSI set $\{\hat{\boldsymbol{h}}_{j,i},\forall j\in{\cal K}\}$ at Tx $i$, problem (\ref{eq.LocalBeam}) can be independently solved by Tx $i$, $\forall i\in{\cal K}$. After that, each Tx $i\in{\cal K}$ broadcasts its $\boldsymbol{t}_i(q+1)$ to the remaining $(K-1)$ Txs.

\item \textbf{Updating inner variable $\boldsymbol{t}$}:
Having obtained $\boldsymbol{t}_i(q+1)$, each Tx-$i$ updates $\boldsymbol{t}$ by solving the problem:
\begin{align}\label{eq.bar_t}
\min_{\boldsymbol{t}}~\frac{c}{2}\sum_{i=1}^{K}||\boldsymbol{E}_i\boldsymbol{t}-\boldsymbol{t}_i(q+1)||^2+\sum_{i=1}^{K}\boldsymbol{\beta}_i^H(q)\boldsymbol{E}_i\boldsymbol{t}.
\end{align}
Note that \eqref{eq.bar_t} is a convex problem and can thus be efficiently solved by convex solvers such as CVX\cite{CVX}.

\item \textbf{Updating dual variables $\boldsymbol{\beta}_i$}:
The dual variables are then updated by the subgradient-based method:
\begin{align}\label{eq.update_dualvariable}
\boldsymbol{\nu}_i(q+1) = \boldsymbol{\nu}_i(q)+c\big(\boldsymbol{E}_i\boldsymbol{t}(q+1)-\boldsymbol{t}_i(q+1)\big),~~\forall i\in{\cal K}.
\end{align}
Set $q=q+1$, and repeat the above updating process until a predefined convergence criterion is satisfied.
\end{itemize}

\begin{remark}
For the proposed ADMM decentralized algorithm, the following comments are in order:
\begin{itemize}
\item The introduction of local vectors $\boldsymbol{t}_i$, $\forall i\in{\cal K}$, is a key ingredient to enable a decentralized operation, i.e., leading to the $K$ subproblems each given by \eqref{eq.LocalBeam}.
\item In order to approach the solution of original problem, a ``consensus'' step is performed in \eqref{eq.bar_t} per Tx-$i$ such that the ``public'' vector $\boldsymbol{t}$ is consistent with all local vectors $\boldsymbol{t}_i$, $\forall i\in{\cal K}$. Such an update of inner variables follows the Gauss-Seidel method\cite{BerBook89}.
\item Together with the subgradient-based update of the dual variables, an ADMM algorithm is complete, where only the information of local vectors $\{\boldsymbol{t}_i\}$ needs to be exchanged among the Txs per iteration.
\item The proposed approach is well suited to the typical interference channel setups where there does not exist a central control unit as well as a backhaul link connecting the distributed Txs.

\end{itemize}
\end{remark}

It is worth noting that with the introduction of penalty terms $({c}/{2})\sum_{i=1}^{K}||\boldsymbol{E}_i\boldsymbol{t}-\boldsymbol{t}_i||^2$, the objective functions in \eqref{eq.LocalBeam} and \eqref{eq.bar_t} become bounded below and strictly convex so that these subproblems always have a unique solution. Following from \cite[Proposition 4.2]{BerBook89} and the convergence condition in Section IV.A, the convergence of the proposed decentralized scheme is then guaranteed, as formally stated below.

\begin{proposition}
The iterates $\boldsymbol{t}_i(q)$, $\boldsymbol{t}(q)$, and $\boldsymbol{\nu}_{i}(q)$ in the proposed decentralized algorithm converge to the optimal
primal and dual solutions of \eqref{eq.wc5} as $q \rightarrow\infty$. When the algorithm converges, the returned \{$\boldsymbol{W}_i,\rho_{i}$\} is a globally optimal solution of the SDP relaxation (P2).
\end{proposition}
\begin{remark}
Since the rank-relaxation of (P2) (i.e., dropping the rank-one constraints) is an instance of SDP, general interior-point methods can output its globally optimal solution with computational complexity of ${\cal O}\left(\sqrt{N}\left(K^5N^6+K^6N^4+K^7N^2\right)\right)$ in a centralized manner\cite{Zhao15}.
On the other hand, in the proposed decentralized algorithm, solving the local SDPs \eqref{eq.LocalBeam} and \eqref{eq.bar_t} require computational complexity of ${\cal O}\left(\sqrt{KN}\left(KN^6+K^2N^4+K^3N^2\right)\right)$ and ${\cal O}\left(K^{6.5}\right)$, respectively, and updating dual variables in \eqref{eq.update_dualvariable} requires a linear complexity ${\cal O}(K)$ per Tx-$i$ per iteration $q$. Due to the fast convergence of ADMM, typically less than $50$ iterations would be required to achieve the global optimum of (P2), as corroborated by our simulations in the sequel. Note that the local problems \eqref{eq.LocalBeam} and \eqref{eq.bar_t} for each Tx can be solved in a parallel manner. The running time of the proposed decentralized algorithm is thus affordable.
\end{remark}

\section{Numerical Results}\label{sec:numerical}
In this section, we provide extensive numerical results to gauge the performance of the proposed ADMM based algorithm for decentralized robust BFPS design. In the simulations, we assume that $\hat{\boldsymbol{h}}_{i,j}\sim \mathcal{CN}(\boldsymbol{0},\boldsymbol{I})$ and
set $\boldsymbol{B}_{i,j}=\epsilon^{-2}\boldsymbol{I}$, $\forall i,j\in{\cal K}$. For simplicity, let $\gamma_i = \gamma$ and $\eta_i=\eta$, $\forall i\in{\cal K}$. The noise variances are set as $\sigma_i^2=-70$ dBm and $\delta_i^2=-50$ dBm, $\forall i\in{\cal K}$; the EH efficiency is set as $\zeta_i=0.25$, $\forall i\in{\cal K}$. All results in the simulations are averaged over 2000 independent channel realizations, and 500 randomizations are generated in the Gaussian randomization
procedure for the relaxed (P2).

\begin{figure}
\centering
\includegraphics[width=3.5in]{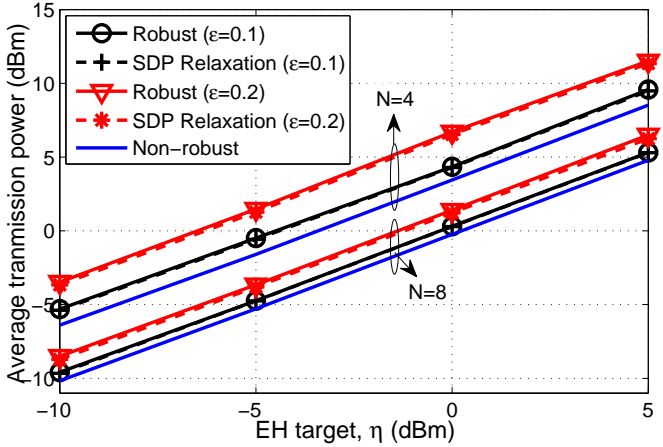}
\caption{The average transmission power versus EH target $\eta$ with $K=2$ and SINR target $\gamma=4$ dB.}\label{fig:power_vs_EH}
\end{figure}

Fig. \ref{fig:power_vs_EH} shows the average transmission power versus the EH target $\eta$ with $K=2$ for different $N$ and different CSI uncertainty bounds.
For comparison, we include the performance of the non-robust BFPS design~\cite{Tim14}, which is based on the estimated channels $\{\hat{\boldsymbol{h}}_{i,j}\}$. In our numerical tests, we observe that most (more than 80\%) of time the SDP relaxation yields rank-one solutions. As a result, the proposed robust BFPS scheme achieves almost the same transmission power as its SDP relaxation. In addition, it is observed that the proposed robust scheme requires slightly more transmission power than the non-robust one in~\cite{Tim14}. This is expected, since the robust scheme accounts for the channel errors when meeting the requirements. As expected, Fig.~\ref{fig:power_vs_EH} also shows more transmission power is required in larger CSI uncertainty cases. When the transmit-antenna number $N$ at each Tx increases from 4 to 8, Fig.~\ref{fig:power_vs_EH} shows that the  transmission power significantly decreases. This demonstrates the benefit by employing more transmit antennas in MISO SWIPT interference channels.

\begin{figure}
	\centering
	\includegraphics[width=3.3in]{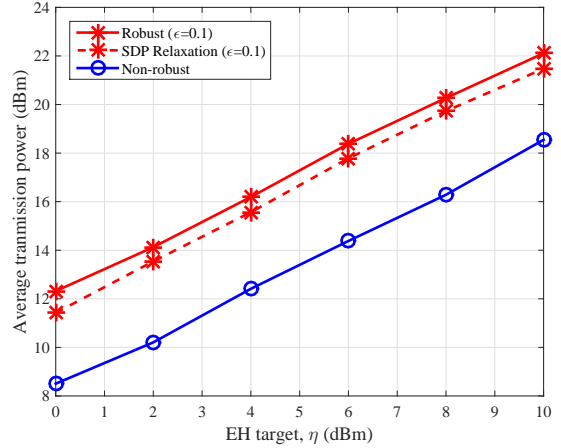}
	\caption{The average transmission power versus EH target $\eta$ with $N=4$, $K=4$, and SINR target $\gamma=5$ dB.}\label{fig:vs_EH_K4}
\end{figure}

Fig. \ref{fig:vs_EH_K4} shows the average transmission power versus the EH target $\eta$ with $N=4$, $K=4$, and the SINR target $\gamma=5$ dB. The channel uncertainty bound is set as $\epsilon=0.1$. It is observed that the proposed robust scheme performs inferior to the SDP relaxation one. This implies that the SDP relaxation cannot always yields rank-one solutions in this setup. Again, it is also observed that the proposed scheme requires slightly more transmission power than the non-robust one in~\cite{Tim14}, since the robust scheme accounts for the channel uncertainties.

\begin{figure}
\centering
\includegraphics[width=3.1in]{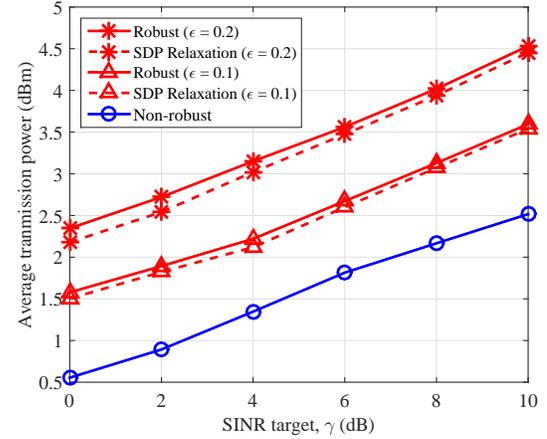}\\
\caption{The average transmission power versus SINR target $\gamma$ with $N=4$, $K=2$, and EH target $\eta=-2$ dBm.}\label{fig:power_vs_SINR1}
\end{figure}

\begin{figure}
	\centering
	\includegraphics[width=3.28in]{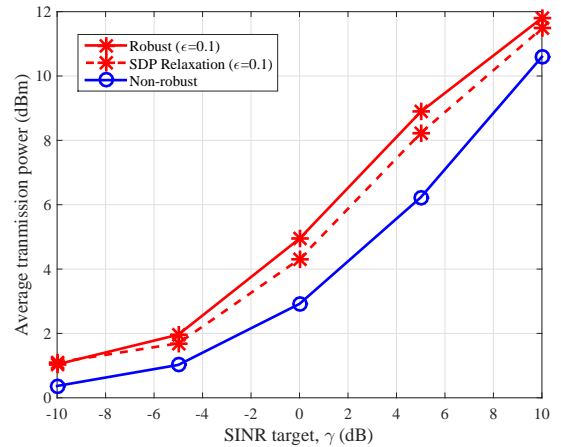}
	\caption{The average transmission power versus SINR target $\gamma$ with $N=4$, $K=4$, and EH target $\eta=-2$ dBm.}\label{fig:vs_SINR_K4}
\end{figure}

Fig.~\ref{fig:power_vs_SINR1} shows the average transmission power versus the SINR target $\gamma$ with $N=4$ and $K=2$ for different CSI uncertainty bounds. The EH target per Rx is set as $\eta=-2$ dBm. As in Fig.~\ref{fig:power_vs_EH}, it is observed that the proposed robust scheme requires more transmission power than the non-robust counterpart, and more transmission power is required in larger CSI uncertainty cases. Since the SDP relaxations cannot always yield rank-one solutions under this setup, a performance gap in terms of transmission power exists between the proposed robust and the SDP relaxation schemes, especially at small SINR targets. Fig.~\ref{fig:vs_SINR_K4} shows the average transmission power versus the SINR target $\gamma$ with $N=4$, $K=4$, and the EH target $\eta=-2$ dBm. The channel uncertainty bound is set as $\epsilon=0.1$. Compared with Fig.~\ref{fig:power_vs_SINR1}, a similar observation is obtained and significantly more power is required for a SWIPT interference channel with a larger user-pair number $K$.

\begin{figure}
\centering
\subfigure{\includegraphics[width=3.5in]{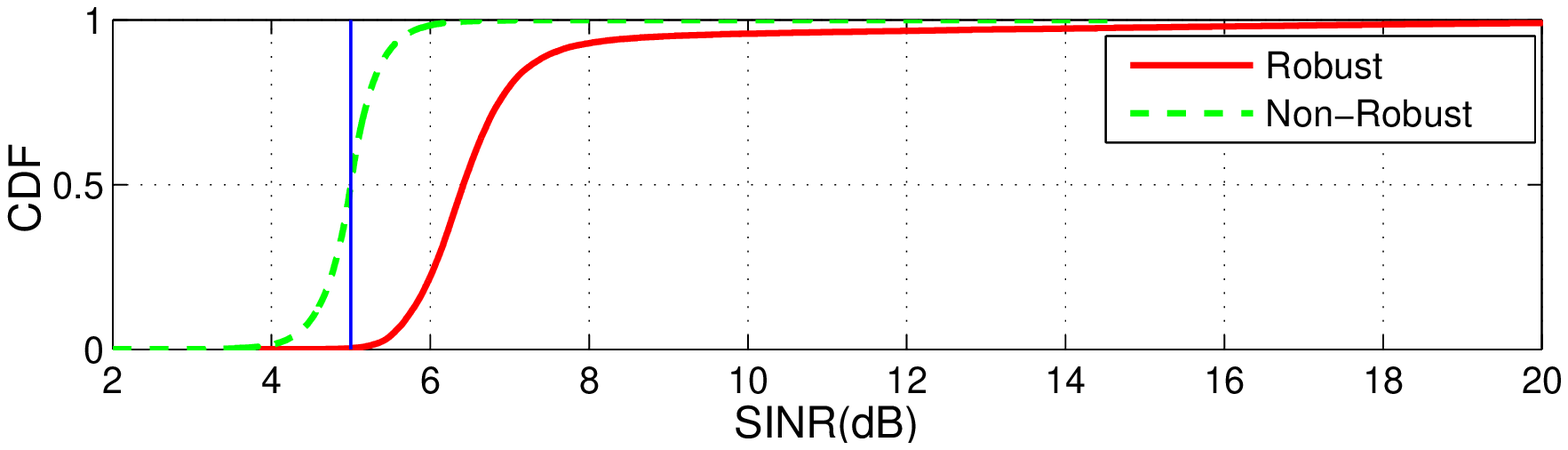}}\\
\subfigure{\includegraphics[width=3.5in]{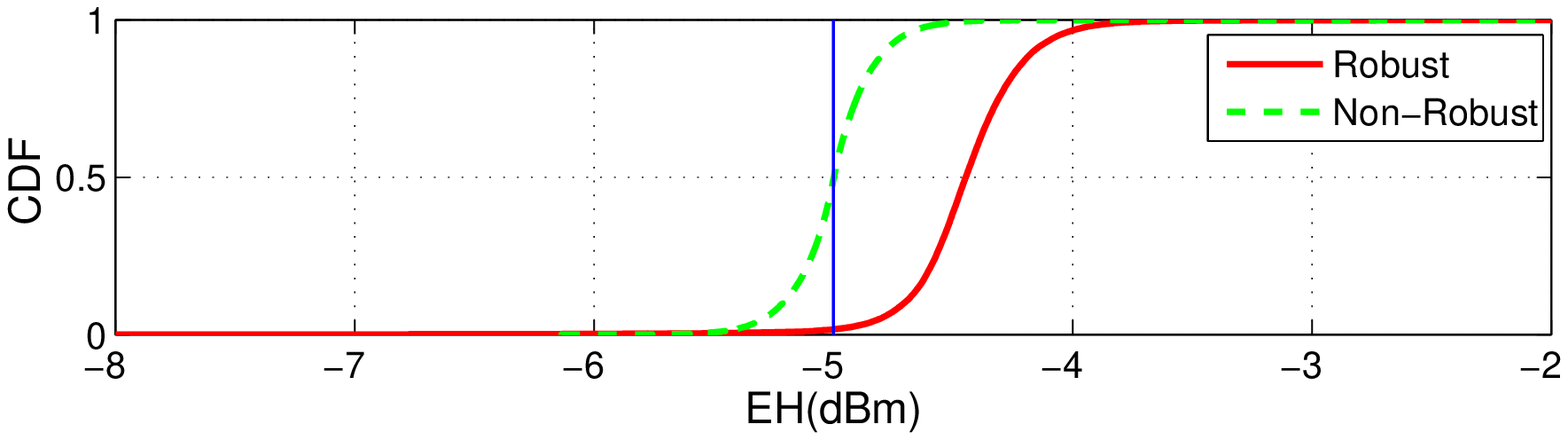}}\\
\caption{CDFs of achieved EH and SINR for $K=2$, $N=4$, the EH target is $\eta = -5$ dBm, and the SINR target $\gamma=5$ dB.}\label{fig:cdf}
\end{figure}

Fig.~\ref{fig:cdf} shows the cumulative distribution functions (CDFs) of the achieved SINR and EH per Rx. It is observed that the proposed robust scheme satisfies the prescribed targets, while the non-robust one only meets these targets about $50\%$ of the time. This implies that the less transmission power of non-robust scheme in Figs.~\ref{fig:power_vs_EH}--\ref{fig:vs_SINR_K4}, is at the price of severely violating the SINR and EH constraints.

\begin{figure}
\centering
\includegraphics[width=3.5in]{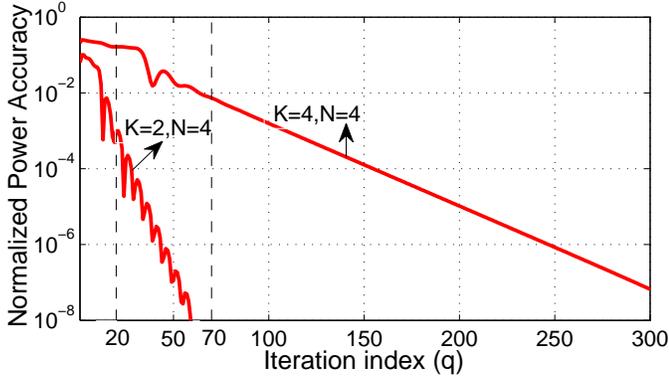}\\
\caption{Convergence curves of the proposed ADMM decentralized algorithm for $\epsilon=0.1$, EH target $\eta = -5$ dBm, and SINR target $\gamma = 5$ dB.}\label{fig:admm}
\end{figure}

Fig.~\ref{fig:admm} shows the convergence performance of the proposed ADMM decentralized algorithm and the vertical axis is the normalized power accuracy defined by $\Delta P \triangleq {|P(q)-P^*|}/{P^*}$, where $P(q)=\sum_{i=1}^K{p_i(q)}$ is the total transmission power at iteration $q$ and $P^*$ denotes the optimal value of problem (P2). The penalty parameter is set as $c=1$. When $K = 2$ and $N = 4$, Fig.~\ref{fig:admm} indicates that the proposed decentralized algorithm  yields a solution with $\Delta P \leq$ 0.01 within 20 iterations. When the user number increases from $K=2$ to $K=4$, it is shown that about 70 iterations are needed for achieving $\Delta P \leq 0.01$. The results in Fig.~\ref{fig:admm} demonstrate the fast convergence of the proposed ADMM decentralized algorithm.

\section{Concluding Remarks}\label{sec:conclusion}

This paper investigated the decentralized robust BFPS design for the $K$-user MISO SWIPT interference channel based on ADMM. By transforming the infinitely many worst-case SINR and EH constraints into compact LMI forms, we showed that the intended BFPS problem can be solved via a convex SDP relaxation in a centralized fashion. Based on the principle of ADMM, we developed a decentralized algorithm capable of computing the optimal transmit beamforming and receive power-splitting schemes with only local CSI and limited information exchange among the Txs. Numerical results demonstrated the merit of the proposed robust approaches.

It is worth noting that the recently proposed EH communication with integration of smart grids presents new challenges in system modeling and design, theoretical analysis, and signal processing\cite{Wang15new,Wang16new2,Chen17,Hu16,Wang16new,Wang17new}, by taking into account two-way energy flows, causal EH profiles, and storage imperfections. Meanwhile, the area of WPT with smart grids provides exciting possibilities to further adapt the energy flows and communication operations. As a future work, a possible extension of this decentralized BFPS work is to consider a SWIPT system powered by smart grids, where the system performance is expected to be further improved in terms of energy efficiency.

%

%
%
%
%
%

\end{document}